\documentclass[aps,prx,twocolumn,superscriptaddress]{revtex4-2}

\usepackage{graphicx}
\usepackage{wrapfig}
\usepackage{setspace}
\usepackage{caption}
\usepackage{etoolbox}
\usepackage{amsmath}
\usepackage{xcolor,soul}
\usepackage{float}

\makeatletter
\patchcmd{\frontmatter@abstract@produce}
{\vskip200\p@\@plus1fil
	\penalty-200\relax
	\vskip-200\p@\@plus-1fil}
{}
{}
{}
\makeatother

\begin{document}

\title{Director--layer dynamics in the antiferroelectric smectic-$Z_A$ phase of a ferroelectric nematic liquid crystal}
	
\author{Arjun Ghimire}
\affiliation{Department of Physics, Kent State University, Kent OH 44242}

\author{Bijaya Basnet}
\affiliation{Materials Science Graduate Program and Advanced Materials and Liquid Crystal Institute, Kent State University, Kent, OH 44242}

\author{Hao Wang}
\affiliation{Advanced Materials and Liquid Crystal Institute, Kent State University, Kent, OH 44242}

\author{Parikshit Guragain}
\affiliation{Department of Chemistry, Kent State University, Kent OH 44242}

\author{Alan Baldwin}
\affiliation{Department of Physics, Kent State University, Kent OH 44242}

\author{Robert Twieg}
\affiliation{Department of Chemistry, Kent State University, Kent OH 44242}

\author{Oleg D. Lavrentovich}
\affiliation{Department of Physics, Kent State University, Kent OH 44242}
\affiliation{Materials Science Graduate Program and Advanced Materials and Liquid Crystal Institute, Kent State University, Kent, OH 44242}

\author{James Gleeson}
\affiliation{Department of Physics, Kent State University, Kent OH 44242}

\author{Antal Jakli}
\affiliation{Department of Physics, Kent State University, Kent OH 44242}
\affiliation{Materials Science Graduate Program and Advanced Materials and Liquid Crystal Institute, Kent State University, Kent, OH 44242}

\author{Samuel Sprunt}
\email[]{ssprunt@kent.edu}
\affiliation{Department of Physics, Kent State University, Kent OH 44242}
\affiliation{Materials Science Graduate Program and Advanced Materials and Liquid Crystal Institute, Kent State University, Kent, OH 44242}

\date{\today}
	
\begin{abstract}
A dynamic light scattering study of director-layer fluctuations in the antiferroelectric smectic-$Z_A$ phase of the ferroelectric nematic liquid crystal DIO is reported. The dynamics are consistent with the distinctive feature of the $Z_A$ phase that the smectic layers form parallel to the axis of molecular orientational order (director). A model is developed to describe quantitatively the dispersion of the fluctuation relaxation rates. The model is based on a specialization of the elastic free energy density of the smectic-$C$ phase to the case of $90^\circ$ director tilt, a "first-order" approximation of the viscous stresses by their form for an incompressible uniaxial fluid, and a treatment of the effect of chevron layer structure that develops in planar sample cells due to temperature-dependent layer shrinkage, as documented in previous studies on DIO. From the modeling, the layer compression elastic constant is estimated to be $\sim 100$ times lower in the smectic-$Z_A$ phase than in an ordinary smectic-$A$ liquid crystal. Possible effects of the antiferroelectric layer polarization on the director splay elasticity and viscosity are described. The temperature dependencies of the splay, twist, and bend elastic constants and associated viscosities in the higher temperature nematic phase are also presented.       
\end{abstract}

\maketitle
\newpage

\section{Introduction}
The discovery of highly polar liquid crystals exhibiting a ferroelectric nematic phase (FNLCs) \cite{Nishikawa2017,Mandle2017,Mandle20172,Chen2020} – anticipated a century earlier by Born \cite{Born1916} – is certainly one of the most important recent developments in soft materials science. Since 2020, there has been an explosion in the synthesis and physical characterization of FNLC compounds, many based on the original molecular architectures or mixtures thereof \cite{Mandle2021,Mandle2022,Chen2022,Saha2022,Cruikshank2024,Adaka2024} and others representing distinct new types \cite{Aya2021,Manabe2021,Huang2022,Aya2024}. The ferroelectric phase is typically observed at temperatures below a conventional (paraelectric) nematic, although it may also form via direct transition from the isotropic state \cite{Manabe2021,Perera2023}. The polarization field $\vec{P}$ (average dipole moment density) coincides with the average direction of the molecular long axis – i.e., $\vec{P} \parallel \hat{n}$, where $\hat{n}$ is the nematic director – and the magnitude of $\vec{P}$ reaches values corresponding to nearly saturated polar ordering of the molecular dipoles.

Remarkably, the ferroelectric nematic is only one member of an expanding “realm” of related ferroelectric and antiferroelectric phases \cite{Brown2021,Chen20222,Chen2023,Giesselmann2024,Karcz2024,Nishikawa2024,Strachan2024,Song2024}, including smectic phases exhibiting polar layer structures. Of particular interest is an antiferroelectric phase, first identified in the compound DIO \cite{Nishikawa2017}, that occurs in a temperature range between the uniaxial paraelectric and ferroelectric nematic states. In DIO, resonant carbon K-edge and off-resonant Xray scattering, combined with measurements of polarization current, reveal antiferroelectric order accompanied by a weak mass density wave (period $d_0 = 9$~nm) running normal to $\hat{n}$ and $\vec{P}$, which alternates with period $2 d_0$ and is essentially saturated in each ``slab" making up the layered structure \cite{Chen2023}. These features led to the designation of a new smectic phase, the smectic-$Z_A$, where ``Z" denotes the limit of a smectic-$C$ phase with a $90^\circ$ degree director tilt from the smectic layer normal, and ``A" indicates that the smectic layers are delineated by the reversal of $\vec{P}$. The key features of the smectic-$Z_A$, including $\sim 10$~nm mass density wave, antiferroelectric response to alternating applied electric fields, and textures indicating a chevron layer structure (associated with temperature-dependent layer shrinkage), have also been reported on mixtures of DIO with other polar molecules \cite{Giesselmann2024}.

%, the smectic-$Z_A$ is apparently not the only manifestation of antiferroelectric ordering above the ferroelectric nematic phase. Indeed
However, experimental results on another prototypical FNLC, the compound RM734 \cite{Mandle2017}, and its mixture with ionic liquids \cite{Rupnik2025}, suggest an alternative model for the antiferroelectric phase. In pure RM734, the onset of polar order is preceded by a dramatic decrease with temperature of the elasticity and relaxation rate associated with splay director fluctuations \cite{Mertelj2018}. When doped with ionic additives, a very narrow antiferroelectric phase ($\sim 1^\circ$C wide) below the paraelectric nematic broadens significantly in temperature and exhibits an optical stripe texture, with the stripes running parallel to the average $\vec{P}$ \cite{Rupnik2025}. Both results were explained by an anti-ferroelectric, splay-modulated phase (designated $N_S$), wherein a periodic reversal in sign of director splay ($\vec{\nabla} \cdot \hat{n}$), and accompanying reversal of $\vec{P} \parallel \hat{n}$, enables the splay structure to fill space efficiently. With ionic doping, the accumulation of bound charge due to the polarization splay is screened by free charge, explaining the broadening of the antiferroelectric phase in the doped samples. At the same time, the inverse flexoelectric effect favors splay in the presence of non-zero $\vec{P}$. Very recently, a doubly splay-modulated antiferroelectric phase has been reported on RM734 samples confined in thin planar cells coated with ionic polymers \cite{Ma2025}.   

The two models for the antiferroelectric phase are not necessarily incompatible: Assuming a one-dimensional splay modulation in the $N_S$ model, the reversal in $\vec{\nabla} \cdot \hat{n}$ (and $\vec{P}$) may coincide with a one-dimensional electron density variation proportional to $P^2$, as suggested in ref.~\cite{Rupnik2025}. Additionally, the splay modulation may develop on a suboptical scale. Further, if $\vec{P}$ is large, splay would tend to be expelled due to high electrostatic energy cost. The combination of these characteristics would imply a smectic layer structure with $\hat{n}$ essentially parallel to the layer planes and $\vec{P}$ reversing between them, matching characteristics of the smectic-$Z_A$ structure.

%One of these, dubbed the smectic-$Z_A$ phase, was first identified \cite{Chen2023} in the compound DIO \cite{Nishikawa2017} in a temperature range between uniaxial paraelectric and ferroelectric nematic phases. It is characterized by an antiferroelectric stacking of polarized fluid layers, with the director and alternating layer polarization fields oriented {\it parallel} to the layer planes. These features were determined \cite{Chen2023,Giesselmann2024} from detailed optical textural studies and polarization current measurements under alternating applied electric fields. Resonant carbon K-edge and off-resonant Xray scattering \cite{Chen2023} definitively established the antiferroelectric stacking (with period $2 d_0 \approx 18$~nm) and revealed a weak smectic mass density wave running normal to the polarized layers with period $d_0$. 

%In terms of the layer-director structure, one may regard the smectic-$Z_A$ phase as the limit of a smectic-$C$ phase with a $90^\circ$ degree molecular tilt from the layer normal, or as the ``complement" (in terms of director orientation) of the ordinary smectic-$A$ phase. 

The smectic-$Z_A$ and $N_S$ models of the antiferroelectric phase in FNLCs represent new motifs for antiferroelectric ordering in LCs, and it is therefore clearly of interest to explore the nature of the orientational fluctuation modes in this phase. Here we report a dynamic light scattering study of these modes in both the paraelectric nematic and antiferroelectric phases of pure DIO. The results in the antiferroelectric phase are analyzed in terms of the smectic-$Z_A$ model by combining the elastic free energy of a smectic-$C$ phase \cite{Hatwalne1995} in the limit of $90^\circ$ tilt angle with the standard hydrodynamic equations for an incompressible uniaxial fluid. 

Although the allowed biaxiality of the smectic-$Z_A$ phase is neglected in the form taken for the dissipative stress, the dynamical model quantitatively describes the dispersion of the fluctuation modes, provided two additional factors are taken into account. First, in the “bookshelf” layer geometry obtained in sample cells treated for uniform planar alignment of the director, the smectic layer orientation is distorted by temperature-dependent variation of the layer spacing, resulting in a chevron layer structure that was previously documented in DIO and various DIO-containing mixtures \cite{Chen2023,Giesselmann2024}. The chevron structure is characterized by a rotation of the layer normal (by angle $\delta$) away from the normal bookshelf orientation; its magnitude in DIO is significantly smaller than in the classic examples of smectics-$C$ cooled from a higher temperature smectic-$A$ phase \cite{Rieker1987,Clark1988}. Second, splay fluctuations in the polarization field $\vec{P} = P_0 \hat{n}$ store electrostatic energy through the accumulation of polarization charge, which modifies the effective energy density associated with splay and is distinct from the elastic constant for  layer tilt. Apart from the quantitative analysis, our experimental results confirm that the average (equilibrium) orientation of the director is parallel to the layer planes. 

The elastic free energy density we use for our analysis, while qualitatively different from a standard smectic-$A$, contains the usual elastic constants for layer compression and tilt. It assumes only that the average $\hat{n}$ is parallel to the layer planes, and in this respect could apply even in the absence of a true smectic density wave, so long as the antiferroelectric modulation of $\vec{P}$ is treated as a ``pseudo"-layered structure with periodicity small compared to optical wavelengths. Based on the model, we obtain an estimate of the layer compression elastic constant, $B$, in the antiferroelectric/smectic-$Z_A$ phase of DIO that is $\sim 100$ times smaller than in a typical calamitic smectic-$A$. However, as we will describe, this estimate is obtained from the quantity $B \delta^2$ and therefore depends strongly on the value of the chevron angle $\delta$ \cite{Chen2023}.  

Additionally, we compare the pretransitional behavior of splay, twist, and bend elasticities and associated viscosities at the paraelectric nematic to antiferroelectric/smectic-$Z_A$ transition with their behavior previously reported in the FNLC compound RM734 \cite{Mertelj2018}.

Because the antiferroelectric phase in DIO unambiguously exhibits a smectic density wave and a temperature-dependent chevron layer structure, we shall hereafter refer to it as smectic-$Z_A$.
	
\section{Experimental details}

Dynamic light scattering and simultaneous polarizing optical microscopy were performed on a $20~\mu m$ thick layer of purified trans DIO contained between planar optical substrates. The inner substrate surfaces were treated with rubbed polyimide alignment layers, and the cell was assembled with the rubbing directions on each surface oriented parallel to each other (syn-parallel alignment) in order to produce homogeneous planar orientation of the nematic director $\hat{n}$. The outer cell surfaces were coated with a high efficiency antireflection coating designed for the visible range. The sample cell was housed in a temperature-regulated hot stage, which has $\pm 50^\circ$ optical access for scattered light and typical temperature stability of $0.01^\circ$C. The hot stage was mounted on a goniometer that allowed independent control of the incident and scattering angles defined in the horizontal scattering plane and measured relative to the normal to the vertical sample plane. An additional rotation stage enabled variation of the nematic director orientation relative to the scattering plane.

The two scattering geometries used are summarized in Fig.~1. They correspond to the twist-bend mode of the nematic director when $\hat{n}$ lies in the scattering plane ({\it Geometry 1}) or to a combination of pure splay and pure twist director fluctuations when $\hat{n}$ is perpendicular to the scattering plane ({\it Geometry 2}). In both cases, the polarization of normally incident laser light (wave vector $\vec{K}_i$, with $| \vec{K}_i | = K_0 = 2 \pi/532$~nm$^{-1}$) was set perpendicular to $\hat{n}$, and the maximum intensity of the focused beam at the sample was $0.1$~W/mm$^2$. Homodyne time correlation functions of the intensity of depolarized quasi-elastically scattered light (wavevector $\vec{K}_s$, with $| \vec{K}_s | = K_0$) were acquired in each scattering geometry for lab scattering angles ($\theta_s$) ranging from $2.5$ to $50^\circ$. In terms of $\theta_s$, the components of the scattering vector $\vec{Q} = \vec{K}_s - \vec{K}_i$ are
\begin{subequations}
	\begin{align}
	Q_y = 0, Q_z = K_0 \sin \theta_s, Q_x = K_0 \left( n_\parallel \sqrt{1 - \frac{\sin^2 \theta_s}{n_\perp^2}} - n_\perp \right) \\
	\intertext{in {\it Geometry 1} and}
	Q_z = 0, Q_y = K_0 \sin \theta_s, Q_x = K_0 \left( \sqrt{n_\parallel^2 - \sin^2 \theta_s} - n_\perp \right)
	\end{align}
\end{subequations}
in {\it Geometry 2}. Here 
%$k_0 = 2 \pi / \lambda_0$ ($\lambda_0 = 532$~ nm), 
$n_{\perp (\parallel)}$ is the refractive index for polarization perpendicular (parallel) to $\hat{n}_0$, and we have neglected the small optical biaxiality measured in the smectic-$Z_A$ phase \cite{Chen2023}. 

Representative examples of our correlation data in the nematic and smectic-$Z_A$ phases are presented in Figs.~S1 and S2. The values of $n_\perp$ and $n_\parallel$ were measured at $532$~nm wavelength as a function of temperature in both phases using a wedge cell technique \cite{Kumari2024}.

\begin{figure}
	%[h!]
	%\centering
	%\captionsetup{font={small,stretch=0}}
	%\captionsetup{belowskip=-10pt}
	\includegraphics[width=.48\textwidth]{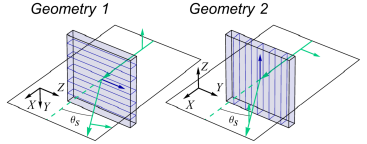}
	\centering
	\caption{Depolarized light scattering geometries used in this study. Blue arrows indicate the equilibrium director orientation ($Z$ direction). The wave vectors of the normally incident laser beam and the scattered light (collected at lab angle $\theta_s = 2.5 - 50^\circ$ from the normal) are shown in green, together with their polarizations. The smectic-$Z_A$ layer planes in the ``bookshelf" layer geometry are rendered with blue outlines.}
\end{figure}

In a typical experimental run, the sample was heated from room temperature to $120^\circ$C in the paraelectric nematic phase, $\sim 36^\circ$C above the transition to the smectic-$Z_A$ phase (at $84^\circ$C) and $\sim 56^\circ$C below the transition to the isotropic phase. (In order to avoid thermal trans-cis isomerization of the DIO molecules, the sample was not heated above $120^\circ$C. Contamination by the cis stereoisomer strongly suppresses the smectic-$Z_A$ phase \cite{Nishikawa2022}). Excellent homogeneous planar alignment of $\hat{n}$ was observed by {\it in situ} polarizing optical microscopy. Previous studies \cite{Chen2023} reveal that in thin cells with parallel surface alignment layers and in the absence of applied bulk fields, the smectic-$Z_A$ layers form in the ``bookshelf" geometry on cooling from the paraelectric phase -- i.e., the layer planes are orthogonal to the cell surfaces at the transition as depicted in Fig.~1. As described below, the ``bookshelf" geometry is confirmed by our light scattering results.  

The depolarized scattered light intensity and its time correlation function were recorded at various temperature on cooling and for various $\theta_s$ at each temperature. The sample texture was examined after each temperature change to confirm that the director alignment remained uniform.

\begin{figure*}
	%[h!]
	%\centering
	%\captionsetup{font={small,stretch=0}}
	%\captionsetup{belowskip=-10pt}
	\includegraphics[width=.95\textwidth]{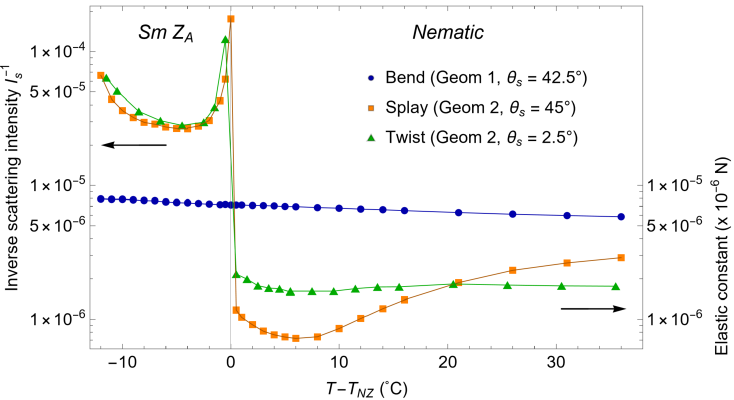}
	\caption{Normalized inverse scattered light intensity (see text) vs temperature contributed by bend, splay, and twist fluctuations of the director in the nematic and smectic-$Z_A$ phases of DIO. For splay and twist, the scale for the elastic constants on the right vertical axis applies to the nematic phase ($T > T_{NZ}$) only. Solid lines are guides to the eye.}
\end{figure*}

\section{Results}

\subsection{Inverse scattered intensity and orientational elasticities}

Fig.~2 displays the temperature dependence of the normalized inverse scattered intensity measured at lab scattering angles where, referring to the coordinates in Fig.~1, $Q \approx Q_z$ in {\it Geometry 1} and $Q \approx Q_y$ or $Q \approx Q_x$ in {\it Geometry 2}. In the nematic phase, the choices of $\theta_s$ isolate scattering from essentially pure bend, splay, and twist director fluctuations, so that the inverse of the depolarized scattering intensity is proportional one of the Frank elastic constants, $K_i$ ($i=1,2,3$ for splay, twist, or bend, respectively). Specifically, $I_{s,i}^{-1} \propto \frac{K_i Q^2}{\Delta \epsilon^2 k_B T G_i}$, where the quantities $G_i = G_i (n_\perp, n_\parallel, \theta_s)$ are calculated geometrical scattering factors, and $\Delta \epsilon = n_\parallel^2 - n_\perp^2$ is the dielectric anisotropy of the uniaxial medium at optical frequency. 

The data plotted in Fig.~2 are the ``normalized" quantities, $I_{s,i}^{-1} \frac{T}{T_{NZ}} (\Delta \epsilon)^2 G_i \left( \frac{K_0^2}{Q^2} \right)$. $\Delta \epsilon$ and the $G_i$ were calculated using the measured $n_\perp$, $n_\parallel$ at each $T$. In the nematic phase, this normalization allows us to obtain absolute values of the $K_i$, provided a ``calibration" value for one elastic constant is available to set the scale. We used the value of $K_1 = 1.2$~pN at $98^\circ$C determined from electric field-induced Freedericksz transition experiments on DIO performed at Kent State. The scaled values of the $K_i$ are presented on the right axis in Fig.~2. 

For temperatures $\gtrsim 20^\circ$C above the nematic to smectic-$Z_A$ transition ($T = T_{NZ}$), the ordering $K_3 > K_1 > K_2$ typical of thermotropic nematics holds. Approaching the transition, the splay constant $K_1$ first gradually decreases (by a factor of $4.5$ from $T-T_{NZ} = 36^\circ$C to $6^\circ$C), then turns around and increases within a $~6^\circ$C range above $T_{NZ}$. This behavior contrasts with the sharper, $\sim 10$ fold decrease in $K_1$ observed over a narrower temperature range above the nematic to ferroelectric nematic transition in the compound RM734 -- a decrease attributed to the coupling of pretransitional fluctuations in the ferroelectric polarization to director splay \cite{Mertelj2018}. However, in DIO the nematic transitions to a smectic phase where the director is parallel to the layer planes; such a layer structure would strongly resist splay with wavevector along the layer normal. One might therefore expect pretranstional fluctuations in the smectic-$Z_A$ order to enhance $K_1$, mitigating the softening effect of flexoelectric coupling. That noted, prior measurements of $K_1$ in DIO (specifically, in the pure trans stereoisomer) \cite{Zhou2022} show a significantly stronger pretransitional decrease than reported here, although the qualitative behavior (including the slight increase in $K_1$ close to the transition) is similar.    

On cooling through the transition, the scattered intensity from bend fluctuations with wavevector along the average director ($Z$ direction in Fig.~1b) is continuous and hardly varies with temperature. On the other hand, the scattering from splay fluctuations with wavevector approximately normal to the layers ($Y$ direction in Fig.~1a) and twist fluctuations with wavevector nearly parallel to the layers ($X$ direction in Fig.~1a) sharply decreases ($I_s^{-1}$ in Fig.~2 increases) by $\sim 100$ times at the transition.

These results are consistent with the distinctive nature of the smectic-$Z_A$ phase -- namely, smectic layers that form parallel to the director field -- as can be understood by the following arguments. As depicted in Fig.~3a (for {\it Geometry 1}), bend distortions, with $\delta \vec{n}$ and $\vec{Q}$ respectively perpendicular and parallel to the layer plane, impose curvature on the layers without requiring changes in layer spacing or incurring defects in layer structure. Therefore bend remains a low energy deformation in the smectic-$Z_A$ phase, and the bend scattering intensity and elastic constant are essentially unaffected by the development of the layer structure. (The same is true for splay scattering from a conventional smectic-$A$, with both $\delta \vec{n}$ and $\vec{Q}$ being parallel to the layer plane in that case.)

\begin{figure}
	%[h!]
	%\centering
	%\captionsetup{font={small,stretch=0}}
	%\captionsetup{belowskip=-10pt}
	\includegraphics[width=.49\textwidth]{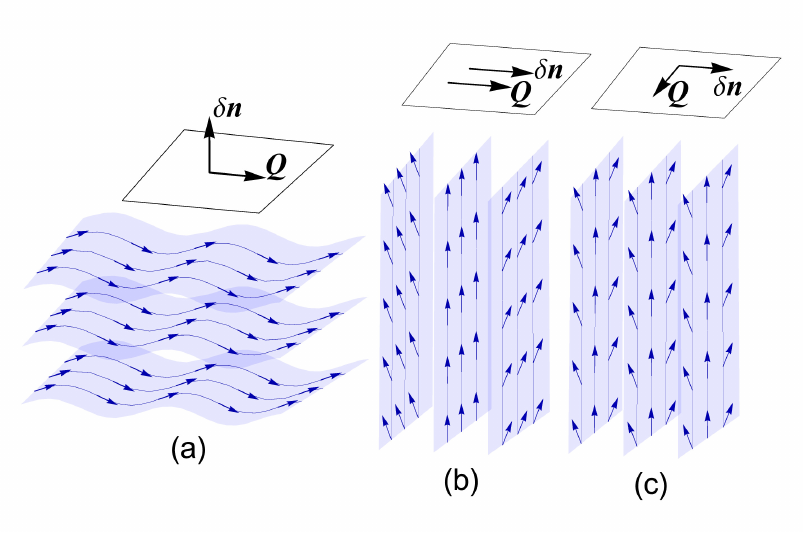}
	\caption{Bend (a), splay (b), and twist (c) director fluctuations ($\delta \vec{n}$) in the smectic-$Z_A$ phase for the case where $\delta \vec{n}$ is normal to the smectic layers and the depolarized scattering cross-section is optimized for the polarizations shown in Fig.~1. The scattering vectors, indicated by $\boldsymbol{Q}$, correspond to those probed in the scattering geometries and at the angles listed in Fig.~2. Black outlines represent the scattering plane. 
	}
\end{figure}

On the other hand, the twist and splay distortions depicted in Figs.~3b and 3c (for {\it Geometry 2}), where $\vec{Q}$ is parallel and perpendicular to the layers, respectively, and $\delta \vec{n}$ is normal to the layers, would incur a significant energy cost in the smectic-$Z_A$ phase due to tilting of the director away from the layer planes or the introduction of defects such as dislocations in the layer structure. Moreover, scattering from twist or splay fluctuations with $\delta \vec{n}$ parallel to the layers in Figs.~3b and 3c (not shown) is strongly suppressed by the selection rules for depolarized scattering.

The arguments above are based on the ``bookshelf" layer geometry, which, as mentioned in sec.~II, is the geometry that develops on cooling into the smectic-$Z_A$ phase in thin, homogeneously aligned samples. The alternative scenario, where the layers form parallel to the substrates, would produce quite different behavior of the light scattering intensity than observed in Fig.~2. In the ``parallel" layer geometry, the layer planes in Fig.~3c would be rotated by $90^\circ$ about the vertical axis, and director twist depicted would involve rotation of the $\hat{n}$ {\it within the layer plane}, leaving the layer orientation unchanged. Then the depolarized scattered intensity from twist fluctuations at small $\theta_s$ should hardly vary at the nematic to smectic-$Z_A$ transition, contrary to the data in Fig.~2.

\begin{figure}
	%[h!]
	%\centering
	%\captionsetup{font={small,stretch=0}}
	%\captionsetup{belowskip=-10pt}
	\includegraphics[width=.48\textwidth]{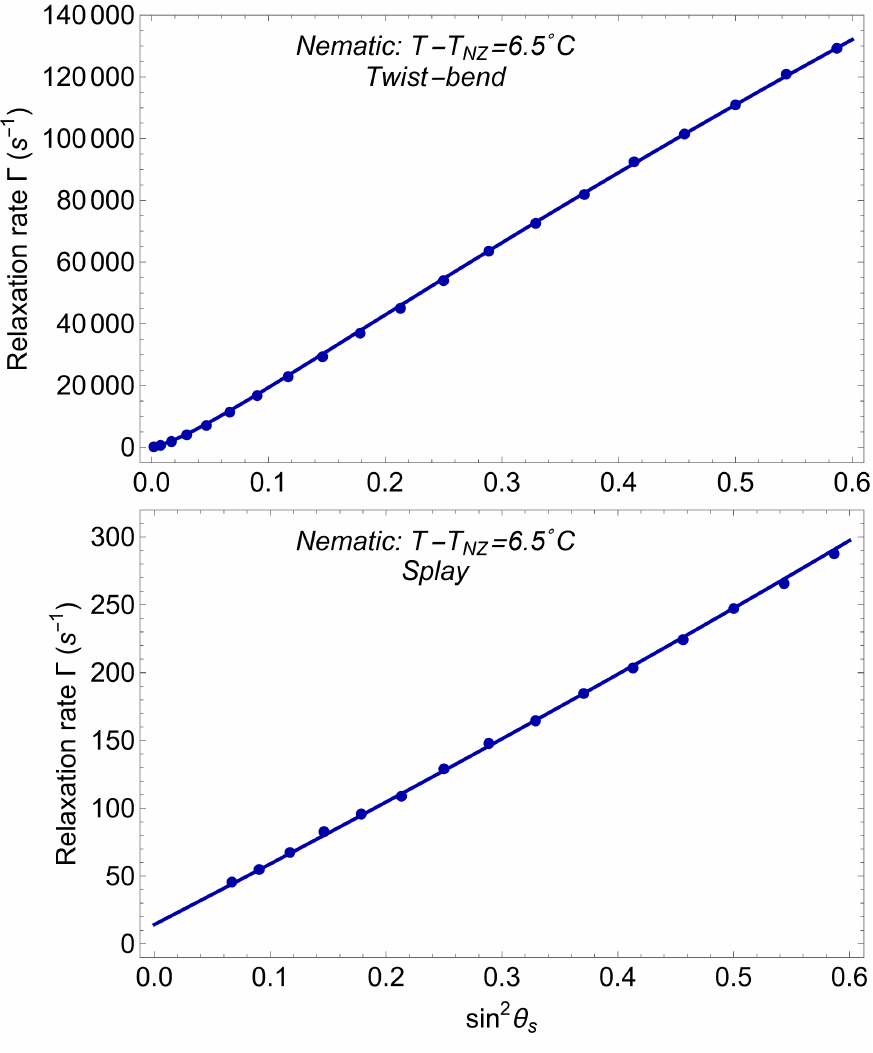}
	\caption{Dispersion of relaxation rates of the twist-bend director mode and the splay component of the splay-bend mode in the nematic phase of DIO. In the latter case, the scattering for $\sin^2 \theta_s \lesssim 0.05$ crosses over to a combination of twist and splay fluctuations, and to nearly pure twist at the smallest $\theta_s$ studied. Solid lines are fits to the expressions for the relaxation rates $\Gamma_{n_2}$ (top panel) and $\Gamma_{n_1}$ (bottom panel) given in the text.}
\end{figure}

The model described in sec.~IVA below shows that twist and splay fluctuations are coupled in the smectic-$Z_A$ phase. Consequently, the scattering recorded from independent modes in the nematic phase at the two scattering angles employed in {\it Geometry 2} becomes scattering predominantly from a single mode (the slower of a pair of coupled modes) at these angles. This in part explains the similar temperature dependence of $I_s^{-1}$ observed for splay/twist fluctuations below $T_{NZ}$ (Fig.~2), as we will discuss further in sec.~IVB. 

\subsection{Relaxation rates and orientational viscosities}

Relaxation rates of the director fluctuations in the nematic phase and director-layer fluctuations in the smectic-$Z_A$ phase of DIO were obtained from fitting the measured time correlation functions of the scattered intensity to single or double exponential decays -- the latter in cases where two well-separated decays were manifest in the data (see examples in Fig.~S1).
 
\begin{figure*}
	%[h!]
	%\centering
	%\captionsetup{font={small,stretch=0}}
	%\captionsetup{belowskip=-10pt}
	\includegraphics[width=.9\textwidth]{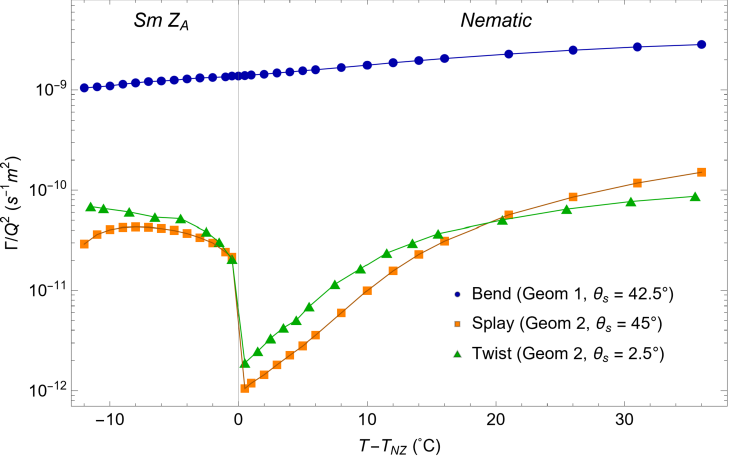}
	\caption{Ratio of relaxation rates to scattering wavenumber squared vs temperature measured in the same scattering geometries as in Fig.~2. In the nematic phase, the data correspond to pure bend, splay, and twist director fluctuations. Solid lines are guides to the eye.}
\end{figure*}

Fig.~4 shows the dependence of the relaxation rates ($\Gamma$) extracted from the fits on the parameter $\sin \theta_s$ at $T-T_{NZ} = 6.5^\circ$C in the nematic phase. The solid lines are fits to the standard expressions for the dependence of $\Gamma$ on $\vec{Q}$ (dispersion of $\Gamma$) in a uniaxial nematic,
\begin{equation*}
\Gamma_{n_2} = \frac{K_2 Q_x^2 + K_3 Q_z^2}{\eta_{twist}-\frac{\alpha_2^2 Q_z^2}{\eta_c Q_z^2 + \eta_a Q_x^2}}
\end{equation*}
for the twist-bend director mode ({\it Geometry 1}), and
\begin{equation*}
\Gamma_{n_1} = \frac{K_1 \left(Q_x^2 + Q_y^2 \right)}{\eta_{splay}}
\end{equation*}
for splay fluctuations ({\it Geometry 2}). Here the labels $n_1$ and $n_2$ refer to the splay-bend and twist-bend normal director modes of a uniaxial nematic. The relation between the components of $\vec{Q}$ and $\sin \theta_s$ is given by Eq.~(1).

The viscosity coefficients $\eta_{twist}$, $\eta_{splay} = \eta_{twist} - \frac{\alpha_3^2}{\eta_b}$ and the Miesowicz viscosities $\eta_a$, $\eta_b$, $\eta_c$ are various combinations of the six Leslie coefficients $\{\alpha_i\}$ ($i=1-6$) characterizing the dissipation in a uniaxial fluid \cite{Leslie,deGennes}. In the fit for $\Gamma_{n_2}$ in the top panel of Fig.~4, three parameters were varied (with $K_2/K_3$ fixed to the value obtained from the data in Fig.~2); however, the fit is dominated by a linear dependence on $Q_z^2 \propto \sin^2 \theta_s$ because $Q_z^2 \gg Q_x^2$ except at small $\theta_s$. $\Gamma_{n_1}$ (bottom panel) was fit with one variable parameter, the ratio $\frac{K_1}{\eta_{splay}}$. The results in Fig.~4 are ``textbook" examples for a uniaxial nematic.

Fig.~5 plots the temperature dependent ratios of relaxation rate to scattering wavenumber for the same scattering geometries as in Fig.~2. In the nematic phase, these correspond to the elasticity to viscosity ratios, $\frac{K_1}{\eta_{splay}}$, $\frac{K_2}{\eta_{twist}}$, and $\frac{K_3}{\eta_{bend}}$, where the bend viscosity is given by $\eta_{bend} = \eta_{twist}-\frac{\alpha_2^2}{\eta_c}$. 

The relaxation rate for bend decreases slowly and continuously through the $T_{NZ}$, again indicating that the layer structure has no impact on bend fluctuations, which is expected if the equilibrium director is parallel to the layers. On the other hand, the relaxation rates for splay and twist decrease by nearly two orders of magnitude as $T \rightarrow T_{NZ}$ and then abruptly increase at $T_{NZ}$, rising in the smectic-$Z_A$ roughly back to their nematic levels. The interpretation of this increase is complicated both by the layer structure, and, as we will argue in sec.~IV below, by the potential effect of the slab polarization on the ratio $\frac{K_1}{\eta_{splay}}$.    
\begin{figure}
	%[h!]
	%\centering
	%\captionsetup{font={small,stretch=0}}
	%\captionsetup{belowskip=-10pt}
	\includegraphics[width=.48\textwidth]{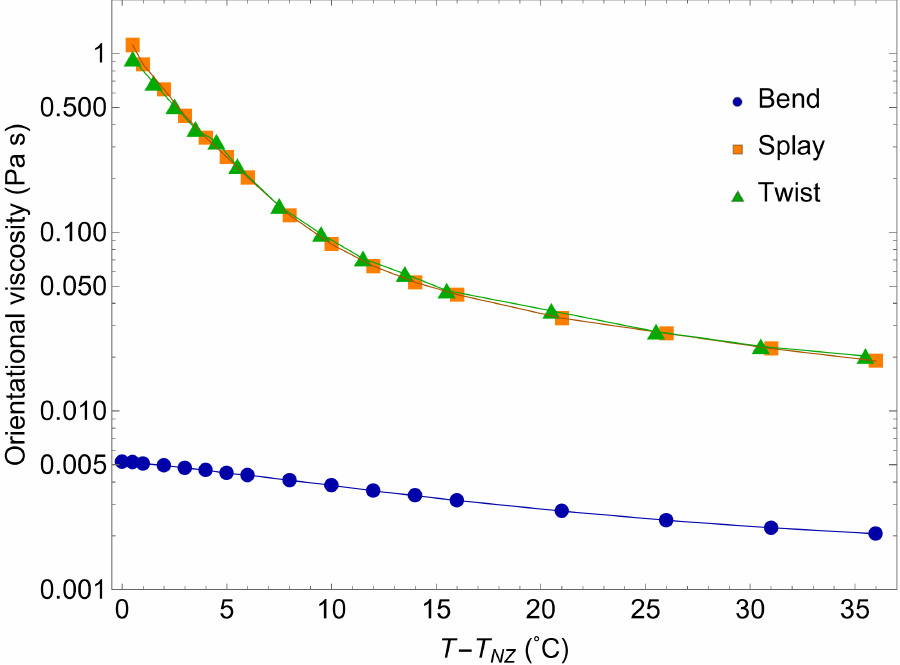}
	\caption{Orientational viscosities $\eta_{bend}$, $\eta_{splay}$, $\eta_{twist}$ vs temperature in the nematic phase of DIO. Solid lines are guides to the eye.}
\end{figure}
\begin{figure*}
	%[h!]
	%\centering
	%\captionsetup{font={small,stretch=0}}
	%\captionsetup{belowskip=-10pt}
	\includegraphics[width=.98\textwidth]{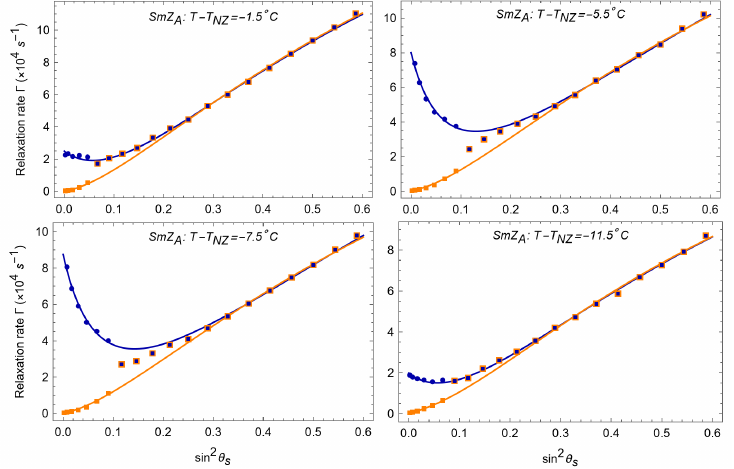}
	\caption{Dispersion of relaxation rates of overdamped fluctuation modes measured in the smectic-$Z_A$ phase in {\it Geometry 1}. The solid blue lines are fits to Eq.~(8) in the text for the relaxation rate ($\Gamma_{n_2,u}^-$) of the slower of a pair of coupled twist-bend/layer displacement ($n_2$, $u$) modes (blue data points). For $\sin^2 \theta_s \lesssim 0.1$, an additional overdamped mode contributes to the light scattering correlation function, with a much slower relaxation rate (orange data points). As discussed in sec.~IVB, we attribute this mode to splay-bend fluctuations within the smectic-$Z_A$ layers. For large $\sin^2 \theta_s$, its relaxation rate is dominated by bend and should be nearly degenerate with $\Gamma_{n_2,u}^-$. The potential degeneracy is indicated by the orange borders surrounding the blue points (in the range of $\sin \theta_s$ where the light scattering correlation is described by a single exponential decay). The orange line represents a fit to an expression for $\Gamma_{n_1}$ based on Eq.~(4) in the text.}
\end{figure*}

The temperature dependence of the orientational viscosities $\eta_{splay}$, $\eta_{twist}$, and $\eta_{bend}$ in the nematic phase, calculated from the results in Figs. 4 and 5, is presented in Fig.~6. The viscosities for splay and twist are essentially equal and increase by an order of magnitude within $\sim 10^\circ$C of the nematic to smectic-$Z_A$ transition. A similar increase is observed within $\sim 1^\circ$C of the para- to ferroelectric nematic transition in RM734 \cite{Mertelj2018}. In contrast, the bend viscosity increases only weakly as $T_{NZ}$ is approached, and remains significantly lower than the other two viscosities.

The dispersion of the relaxation rates of director fluctuations in the smectic-$Z_A$ phase differs qualitatively from the results in Fig.~4 for the nematic. As demonstrated in Fig.~7, the nematic twist-bend mode (recorded in {\it Geometry 1}) splits into two distinct modes for $\sin^2 \theta_s \lesssim 0.1$, with relaxation rates that have opposing dependencies on $\sin^2 \theta_s$. The splitting varies significantly with temperature, and is most pronounced in the middle of the smectic-$Z_A$ range. In this range the faster mode increases sharply with decreasing $\theta_s$ and intercepts the vertical axis ($\theta_s=0$) at a large value. The intercept decreases substantially at temperatures close to the transitions to the paraelectric and ferroelectric nematic phases, where the layer structure disappears. On the other hand, the slower mode slows down continuously as $\theta_s \rightarrow 0$. For $\sin^2 \theta_s \gtrsim 0.1$, the two modes apparently coalesce into a single overdamped mode, whose time correlation function is described by single exponential decay and whose relaxation rate and dependence on $\sin^2 \theta_s$ are similar to the bend mode measured in the nematic phase for the same range of scattering angle.

\begin{figure*}
	%[h!]
	%\centering
	%\captionsetup{font={small,stretch=0}}
	%\captionsetup{belowskip=-10pt}
	\includegraphics[width=.98\textwidth]{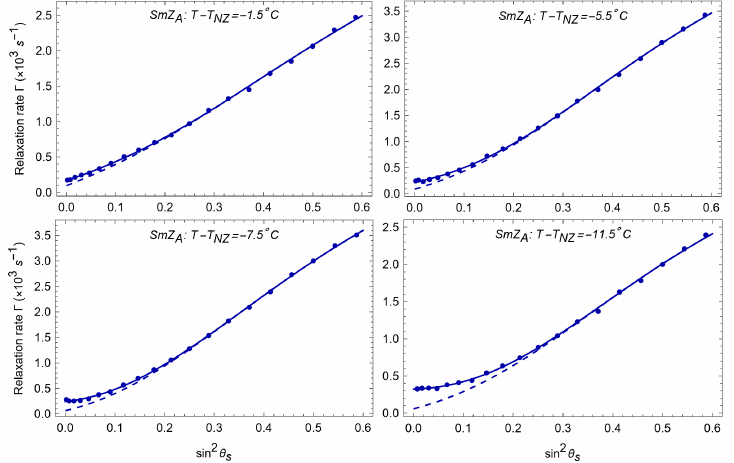}
	\caption{Dispersion of the relaxation rates of the overdamped fluctuation mode measured in the smectic-$Z_A$ phase in {\it Geometry 2}. The dashed blue lines fits to Eq.~(9) in the text for the relaxation of the slower of a pair of coupled splay and twist director modes. The solid blue lines are fits to Eq.~(10) using an effective splay energy density $K_1^{qff} Q_\perp^2$ that incorporates energy stored in the bound charge density produced by splay of the layer polarization.}   
\end{figure*}

In {\it Geometry 2}, where splay fluctuations dominate the scattering in the nematic phase, a single overdamped mode is detected below $T_{NZ}$ over the full range of $\theta_s$. However, its relaxation rate vs $\sin^2 \theta_s$ is not strictly linear (Fig.~8), as expected (and observed in Fig.~4) for the splay component of the splay-bend director mode. Also, as already evidenced in Fig.~5, the relaxation rate is substantially higher than in the nematic phase at the same $\theta_s$.

The distinct features associated with the dispersion of relaxation rates in the smectic-$Z_A$ phase must be a consequence of the layer structure and potentially also of the alternating polarization field in the slabs that define the layers. In the next section, we develop a quantitative model for the dispersion of the relaxation rates that describes the data in Figs.~7 and 8. Crucial to this model is an additional experimental observation -- namely, results from prior studies on DIO \cite{Chen2023,Giesselmann2024} that demonstrate a weak chevron distortion of the layer structure, which develops due to layer shrinkage on cooling within the smectic-$Z_A$ phase.  

\section{Discussion}

In order to model the behavior of the relaxation rates in Figs.~7 and 8, we focus on a simplified model for the hydrodynamics of the smectic-$Z_A$ phase that contains a tractable number of parameters. In particular, we combine an expression for the elastic free energy of a smectic-$C$, developed by Hatwalne and Lubensky (HL) \cite{Hatwalne1995} and extended to the case of $90^\circ$ director tilt, with the hydrodynamic equations for an incompressible, uniaxial fluid having negligible inertia (appropriate for the analysis of overdamped fluctuations). We assume that the impact of biaxiality on the form of the dissipative stress can be neglected. While this assumption has not been validated by direct measurements of viscous coefficients, it nevertheless allows a fairly good quantitative account of our light scattering results.   

\subsection{Elastic free energy density and dynamical equations for the smectic-$Z_A$ phase}

HL present a rotationally-invariant expansion of the elastic free energy density of a smectic-$C$ phase in terms of small deviations ($\delta \hat{n}$) of the director from its equilibrium orientation ($\hat{n}_0$) and small displacements ($u$) of the smectic layers from their equilibrium positions. The elastic energy density is the sum of the usual Frank elastic energy density of a uniaxial nematic (presumed to be the higher temperature phase), a contribution, $f_{Sm}$, associated with the smectic-$C$ layering, and a term $f_{\hat{N}}$ that is quadratic in rotationally invariant derivatives of the layer orientation specified by layer normal $\hat{N}$. 

To quadratic order in variations, $\delta q_s = -\vec{q}_0 \cdot \nabla u$, of the smectic layering wavenumber $q_s$ from its equilibrium value $q_0$ and variations $\delta \Phi$ of the director tilt angle $\Phi$ from its equilibrium value $\Phi_0$, HL obtain
\begin{equation}
	f_{Sm} = \frac{1}{2} B \left[ \hat{q}_0 \cdot \nabla u +  c \delta \Phi \right]^2 + \frac{1}{2} D ( \delta \Phi )^2 . \nonumber
\end{equation}
The coefficients $B = \frac{1}{q_0^2} \left( {\partial^2 f_{Sm} (q_s,\cos\Phi_0)} \over {\partial q_s^2} \right)_{q_s=q_0(\cos \Phi_0)}$ and $D = \left( {\partial^2 f_{Sm} (q_0(\cos \Phi),\cos\Phi)} \over {\partial \Phi^2} \right)_{\Phi=\Phi_0}$ are smectic elastic constants that correspond to layer compression and layer tilt relative to the director, respectively; they vanish in the absence of smectic order. The parameter $c$ is given by $c = \frac{1}{q_0} \left( {\partial q_0(\cos \Phi)} \over {\partial \Phi} \right)_{\Phi=\Phi_0}$ where $q_0 (\cos \Phi)$ is the value of $q_s$ that minimizes $f_{Sm}$ for a given value of $\cos \Phi$. (Thus, $q_0 (\cos \Phi_0) \equiv q_0$ corresponds to the global minimum.)

In the smectic-$Z_A$ phase, $\Phi_0 = \frac{\pi}{2}$, and we may take $\hat{n}_0 = \hat{z}$ and $\hat{q}_0 = \hat{N}_0 = -\hat{y}$, with the average layer planes being parallel to the $x$-$z$ plane. We distinguish ``layer" coordinates ($x$,$y$,$z$) from the fixed ``lab" coordinates ($X$,$Y$,$Z$) defined in Fig.~1, and used to express the scattering vector $\vec{Q}$ in Eq.~(1), in order to treat the effect of chevron layer structure as described in sec.~IVA below. Then, using $\delta (\cos \Phi) = -\sin \Phi_0 \delta \Phi = \delta (\hat{n} \cdot \hat{N}) = \delta \hat{n} \cdot \hat{N}_0 + \hat{n}_0 \cdot \delta \hat{N}$,  $\delta \hat{n} = n_x \hat{x} + n_y \hat{y}$, and $\delta \hat{N} = N_x \hat{x} + N_z \hat{z} = -\partial_x u \, \hat{x} - \partial_z u \, \hat{z}$  -- all to lowest order in the fluctuating quantities -- we get $\delta \Phi = -n_y + \partial_z u$. Also, in the smectic-$Z_A$ phase, the coefficient $c = 0$ in the expression for $f_{Sm}$ above: By symmetry, $q_0 (\cos (\pi/2+\delta \Phi)) = q_0 (\cos (\pi/2-\delta \Phi))$. Assuming ${\partial q_0(\cos \Phi)} \over {\partial \Phi}$ is continuous, this implies ${{\partial q_0(\cos (\pi/2+\delta \Phi)} \over {\partial \delta \Phi}} = 0$ or $\left( {\partial q_0(\cos \Phi)} \over {\partial \Phi} \right)_{\Phi = \Phi_0} = 0$ for $\Phi_0 = \pi/2$, the equilibrium value of $\Phi$ in the smectic-$Z_A$ phase, and thus $c=0$.

We may use these results to express $f_{Sm}$ in terms of $n_y$ and $u$. Then including the Frank energy density for the uniaxial director and taking the simplest (isotropic) form for $f_{\hat{N}}$, $f_{\hat{N}} = \frac{1}{2} K_u [(\partial_x N_x)^2 +(\partial_y N_y)^2]$, where $K_u$ is an elastic constant for layer curvature, we arrive at the following form for the elastic free energy density $f_Z$ of the smectic-$Z_A$ phase:
\begin{eqnarray}
	f_Z = && \frac{1}{2} K_1 \left( \partial_x n_x + \partial_y n_y \right)^2 + \frac{1}{2} K_2 \left( \partial_x n_y - \partial_y n_x \right)^2 + \nonumber \\ && \frac{1}{2} K_3 \left[ \left( \partial_z n_x \right)^2 + \left( \partial_z n_y \right)^2 \right] + \frac{1}{2} B \left( \partial_y u \right)^2 + \nonumber\\ && \frac{1}{2} D \left( n_y - \partial_z u \right)^2  + \frac{1}{2} K_u \left( \partial_x^2 u + \partial_z^2 u \right)^2 . \nonumber
\end{eqnarray}

$f_Z$ has a form similar to the elastic energy density for an ordinary smectic-$A$ phase, but the term that couples director deformations to gradients in the layer displacement differs significantly. In the smectic-$A$, the coupling is isotropic in the layer plane, whereas in the smectic-$Z_A$ only gradients in $u$ along a single direction in the layer plane (the direction $\hat{n}_0 = \hat{z}$) couple to director distortions (specifically to the component of $\hat{n}$ along the layer normal). This anisotropy of the layer-director coupling has important consequences for the fluctuation modes.
 
Fourier transforming $f_Z$ to $\vec{q}$ space, where $\vec{q}$ is the fluctuation wave vector expressed in the ``layer" coordinates, and defining $q_\perp n_1 = q_x n_x + q_y n_y$, $q_\perp n_2 = q_x n_y - q_y n_x$ give
\begin{eqnarray}
	f_Z = && \frac{1}{2} \left( K_1 q_\perp^2 + K_3 q_z^2 \right) |n_1|^2 +\frac{1}{2} \left( K_2 q_\perp^2 + K_3 q_z^2 \right) |n_2|^2 + \nonumber\\ && \frac{1}{2} B q_y^2 |u|^2 + \frac{1}{2} D \bigg| \frac{q_y}{q_\perp} n_1 + \frac{q_x}{q_\perp} n_2 - i q_z u \bigg|^2 + \nonumber\\ && \frac{1}{2} K_u \left( q_x^2 + q_z^2 \right)^2 |u|^2 .
\end{eqnarray}
where $q_\perp^2 = q_x^2 + q_y^2$.

In the nematic phase, $B = D = K_u = 0$ and Eq.~(2) reduces to the elastic free energy of a uniaxial nematic with normal director modes $n_1$ (splay-bend) and $n_2$ (twist-bend).

Dynamical equations for the fluctuating variables $n_1$, $n_2$, and $u$ may be obtained from the basic hydrodynamic theory for an incompressible smectic (see Appendix). As mentioned previously, we simplify the analysis with two further approximations: First, the dissipative stress is taken as that of a uniaxial fluid -- i.e., we assume that to first order the biaxiality of the smectic-$Z_A$ can be neglected. Second, we assume that permeation of molecules between layers is a process which is weakly coupled to $\delta \vec{n}$ and whose effect can be ignored.

\subsubsection{\underline{Geometry 1: $\hat{n}_0 \parallel$ scattering plane}} 

In the coordinates of Fig.~1, the scattering vector for {\it Geometry 1} lies in the $X$-$Z$ plane, and its $Y$ component vanishes. Then, assuming an ideal bookshelf layer orientation with layer normal along $\hat{Y}$, the fluctuations producing the scattering have $q_y = Q_y = 0$, and the layer compression ($B$) term in Eq.~(2) drops out. As we shall see below, this leads to calculated relaxation rates for the director-layer modes that do not account for the strong increase in measured relaxation rate of faster mode in Fig.~7 as $\sin^2 \theta_s \rightarrow 0$ at temperatures in the middle of the smectic-$Z_A$ range.

The observed dispersion of the faster mode can be modeled accurately if we consider a weak layer reorientation due to a ``chevron" deformation of the ideal bookshelf layer geometry (Fig.~9). The chevron structure develops when the layer spacing contracts with temperature and the layers buckle in order to fill space. X-ray diffraction confirms the layer contraction in the smectic-$Z_A$ phase of DIO, and indicates a maximum rotation of the layer normal (chevron angle) of $\delta \simeq 10^\circ$ in the middle of the $Z_A$ range \cite{Chen2023}. Polarizing microscopy on thin ($\sim 1 - 3~\mu$m) planar-aligned DIO samples \cite{Chen2023,Giesselmann2024} and on our thicker ($20~\mu$m) light scattering cell (see Fig.~S3) reveal textural features in the smectic-$Z_A$ phase characteristic of the chevron layer structure.

\begin{figure}
	%[h!]
	%\centering
	%\captionsetup{font={small,stretch=0}}
	%\captionsetup{belowskip=-10pt}
	%\includegraphics[width=1\textwidth,trim={2.9cm 1.1cm 2.9cm 2.3cm},clip]{fig3.eps}
	\includegraphics[width=.3\textwidth]{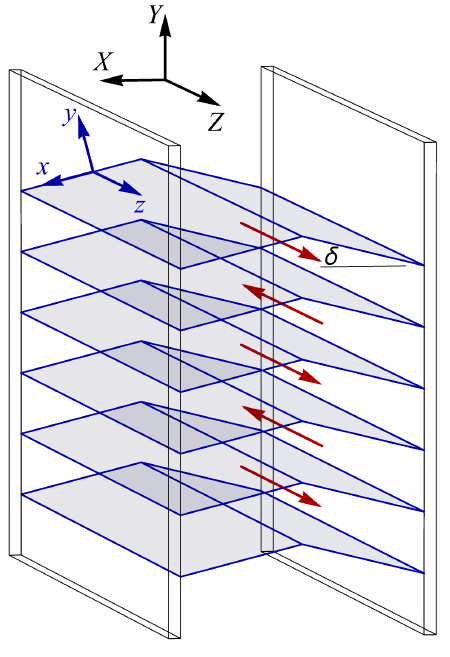}
	\caption{Schematic of chevron layer structure in a thin smectic-$Z_A$ sample confined between parallel substrates. Relative to the lab coordinates (shown in black capitals), the $x$, $y$ axes in the layer coordinates (in blue) are rotated about the $Z$ axis by the chevron angle $\delta$. The molecular director is parallel to the substrates and along the $Z=z$ direction. Red arrows represent the alternating (anti-ferroelectric) layer polarization. The scattering plane in {\it Geometry 1} ({\it 2}) is the $X$-$Z$ ($X$-$Y$) plane (see Fig.~1).}
\end{figure}

As a consequence of the later reorientation, the ``layer" coordinates ($x$,$y$,$z$) in Fig.~9 are rotated from the lab coordinates ($X$,$Y$,$Z$) by $\pm \delta$ around the $z = Z$ axis, and the scattering vector $\vec{Q}$ in the lab has components along the layer normal as well as in the layer plane. This results in the following relation between $\vec{Q}$ and the fluctuation wavevector $\vec{q}$ defined in the ``layer" coordinates,
\begin{eqnarray}
&&q_y = Q_y \cos \delta + Q_x \sin \delta = Q_x \sin \delta \approx Q_x \delta \nonumber\\
&&q_x = Q_x \cos \delta - Q_y \sin \delta = Q_x \cos \delta \approx Q_x \nonumber\\
&&q_\perp = \sqrt{q_x^2 +q_y^2} = \sqrt{Q_x^2 +Q_y^2} = Q_\perp \nonumber\\
&&q_z = Q_z \nonumber ,
\end{eqnarray}
where in the first two lines we put $Q_y = 0$ for {\it Geometry 1} (see Fig.~1) and assumed small $\delta$ in the last expressions on the right. 

When the above substitutions are made in the hydrodynamic equations and in $f_{Z}$ (Eq.~(2)), the dependence on $\delta$ produces additional couplings between the $n_1$, $n_2$, and $u$ that complicate the analysis of the normal modes. A simplification for small $\delta$ sufficient to describe our experimental results is to let $\delta \rightarrow 0$ except in the term of the elastic free energy involving layer compression. This term gives $\frac{1}{2} B q_y^2 |u|^2 = \frac{1}{2} B \delta^2 Q_x^2 |u|^2$ to lowest order in $\delta$. From Eqs.~(A1)-(A4) in the Appendix, we then obtain
\begin{subequations}
\begin{alignat}{2}
\eta_1 (\vec{Q}) \frac{\partial n_1}{\partial t} +& (K_1 Q_x^2 + K_3 Q_z^2) n_1 = 0 \\
\eta_{twist} \frac{\partial n_2}{\partial t} + (D +& K_2 Q_x^2 + K_3 Q_z^2) n_2 \nonumber \\ 
- i D Q_z u +& i \alpha_2 Q_z \frac{\partial u}{\partial t} = 0 \\
\left( Q_x^2 \eta_a + Q_z^2 \eta_c \right) \frac{\partial u}{\partial t} +& \left( B \delta^2 Q_x^2 + D Q_z^2 + K_u Q_\perp^4 \right) u \nonumber \\
+ i D Q_z n_y -& i \alpha_2 Q_z \frac{\partial n_y}{\partial t} = 0
\end{alignat}
\end{subequations}
for the case where $\hat{n}_0$ lies in the scattering plane ({\it Geometry 1}).

In Eqs.~(3), $n_1$ decouples from $n_2$ and $u$, and describes splay-bend director fluctuations within the smectic-$Z_A$ layers. Ignoring the effect of small $\delta$ (to lowest order), the viscosity $\eta_1 (\vec{Q})$ has limiting values $\eta_{splay}$ when $Q_z = 0$ ($\theta_s = 0$) and $\eta_{bend}$ when $Q_z^2 \gg Q_x^2$ (large $\theta_s$). The effect of layer compression appears only in the coupled equations for $n_2$ and $u$ (and only for finite $\delta$).

To calculate the relaxation rates $\Gamma$ of the normal modes, we put $n_1, n_2, u \propto \exp (-\Gamma t)$ in Eqs.~(3), which yields
\begin{equation}
\Gamma = \Gamma_{n_1} = \frac{1}{\eta_1 (\vec{Q})} (K_1 Q_x^2 + K_3 Q_z^2)
\end{equation}
for the relaxation rate of $n_1$, and a quadratic equation
\begin{eqnarray}
&\left[ \eta_{twist} \eta_a Q_x^2 + (\eta_{twist} \eta_c - \alpha_2^2) Q_z^2 \right] \Gamma^2 - \nonumber \\
&\left[ \eta_{twist} (B \delta^2 Q_x^2 + D Q_z^2 + K_u Q_\perp^4) + D (\eta_a Q_x^2 +  \right. \nonumber\\
&\left. (\eta_c + 2 \alpha_2) Q_z^2) + (\eta_a Q_x^2 + \eta_c Q_z^2) (K_2 Q_x^2 + K_3 Q_z^2) \right] \Gamma + \nonumber \\
&D K_3 Q_z^4 + (K_u Q_\perp^4 + B \delta^2 Q_x^2) (D + K_2 Q_x^2) + \nonumber \\
&[K_3 K_u Q_\perp^4 +(D K_2 + B \delta^2 K_3) Q_x^2] Q_z^2 = 0
\end{eqnarray}
whose two roots are relaxation rates $\Gamma_{n_2,u}^+$ and $\Gamma_{n_2,u}^-$ corresponding to the faster and slower normal mode combinations of ($n_2$, $u$), respectively. 

\subsubsection{\underline{Geometry 2: $\hat{n}_0 \perp$ scattering plane}}

In {\it Geometry 2}, $q_z = Q_z = 0$ and the coupling of $n_1$, $n_2$ to layer displacement $u$ drops out in the elastic free energy (Eq.~(2)). Analysis of the hydrodynamic equations with $q_y = Q_y$, $q_x = Q_x$ (to lowest order where $\delta \rightarrow 0$) yields two coupled equations for the fluctuations $n_1$, $n_2$,
\begin{subequations}
\begin{alignat}{2}
\eta_{splay} \frac{\partial n_1}{\partial t} &+ \left(K_1 Q_\perp^2 + D \frac{Q_y^2}{Q_\perp^2} \right) n_1 \nonumber\\
&+ D \frac{Q_x Q_y}{Q_\perp^2} n_2 = 0 \\
\eta_{twist} \frac{\partial n_2}{\partial t} &+ \left(K_2 Q_\perp^2 + D \frac{Q_x^2}{Q_\perp^2} \right) n_2 + D \frac{Q_x Q_y}{Q_\perp^2} n_1 = 0 .
\end{alignat}
\end{subequations}

The secular equation that follows from substituting $n_1, n_2 \propto \exp (-\Gamma t)$ is
\begin{eqnarray}
&\eta_{splay} \eta_{twist} Q_\perp^2 \Gamma^2 - \left[ \eta_{twist} K_1 Q_\perp^4 + \right. \nonumber\\
&\left. \eta_{splay} K_2 Q_\perp^4 + D (\eta_{splay} Q_x^2 + \eta_{twist} Q_y^2) \right] \Gamma \nonumber\\
&\left[ K_1 Q_\perp^2 (K_2 Q_\perp^4 + D Q_x^2) + D K_2 Q_y^2 Q_\perp^2 \right] = 0 ,
\end{eqnarray}
which yields relaxation rates $\Gamma_{n_1,n_2}^+$ and $\Gamma_{n_1,n_2}^-$ for faster and slower normal modes of the coupled fluctuations ($n_1$, $n_2$). When $D \rightarrow 0$, these modes reduce to pure splay and twist modes. 

\subsection{Comparison of the model to experimental results}

First we briefly consider case of a smectic-$Z_A$ sample with an ideal bookshelf layer geometry (i.e., no chevron deformation). For scattering in {\it Geometry 1}, this means not only that $n_1$ decouples from ($n_2$, $u$) in the dynamical equations, but also $n_1$ does not contribute to the scattering cross-section. In the undistorted bookshelf geometry, the scattering and layer planes coincide; for splay-bend fluctuations within the layers ($n_1$ mode), there is no component of the director along the incident light polarization, which is normal to the scattering plane. Then one of the ($n_2$, $u$) modes would have to account for the slow mode and the behavior of its relaxation rate at small $\theta_s$ in Fig.~7. 

In the limit $\theta_s \rightarrow 0$ ($Q_z \rightarrow 0$), the slower ($n_2$, $u$) mode (relaxation rate $\Gamma_{n_2,u}^-$) reduces to pure undulation of the smectic layers, without rotation of the director and, consequently, produces no depolarized light scattering. This contradicts our experimental result that the slower mode scatters intensely as $\theta_s \rightarrow 0$. 

On the other hand, for $\delta = 0$ and small $\theta_s$, the relaxation rate of the faster ($n_2$, $u$) mode increases with $\sin \theta_s$ as $\Gamma_{n_2,u}^+ \approx \frac{D}{\eta_{twist}} + \left( 1- \frac{\alpha_2}{\eta_{twist}} \right)^2 \frac{D}{\eta_a (\Delta n)^2} \sin^2 \theta_s$, assuming $D \gg (K_u,K_2,K_3) K_0^2 (\Delta n)^2$. The latter condition would be necessary to account for the large value of the relaxation rate observed for the faster mode when $\theta_s \rightarrow 0$ in the middle of the smectic-$Z_A$ phase. The predicted increase of $\Gamma_{n_2,u}^+$ with $\sin^2 \theta_s$ at small $\theta_s$ clearly contrasts with the data for faster mode in Fig.~7. Apparently, the model with $\delta = 0$ simply fails to explain our experimental results for {\it Geometry 1}.      

Including the effects of finite $\delta$ resolves these difficulties. When $\delta \neq 0$, the layer planes are canted with respect to the scattering plane (Fig.~9), and splay-bend fluctuations in the layers ($n_1$ mode) produce a component of $\hat{n}$ along the incident light polarization in {\it Geometry 1}. Although this component is small, the $n_1$ fluctuations, which do not involve layer tilt or compression, have a large intrinsic amplitude (low energy) at small $\theta_s$. Additionally, the layer compression term, contributing to Eq.~(3c) for finite $\delta$, profoundly alters the dispersion of the ($n_2$, $u$) modes at small $\theta_s$, leading to an expression for the relaxation rate $\Gamma_{n_2,u}^-$ that accurately describes the data for the upper branch of relaxation rates in Fig.~7. For large $\theta_s$, where $Q_z \gg Q_x$, the $n_1$ mode and the slower ($n_2$, $u$) mode both represent mainly bend director fluctuations, and are nearly degenerate if $K_3 \gg K_u$. This explains why only a single mode is resolved in the light scattering correlation functions at large $\theta_s$. 

A quantitative analysis of the data in Figs.~7 and 8, based on the full expressions for the relaxation rates given by Eqs.~(4), (5), and (7) involves a large number of material parameters, even when the biaxiality of the smectic-$Z_A$ phase is neglected. However, the analysis may be simplified by assuming $D \gg (B \delta^2 , K_i q^2)$ for small $\delta$ and for $q$ in the optical range. In that case, the solution to Eq.~(5) for the slower ($n_2$, $u$) mode in {\it Geometry 1} may be approximated by
\begin{equation}
\Gamma_{n_2,u}^- = \frac{B \delta^2 Q_x^2 + K_u (Q_x^2 + Q_z^2)^2 + K_2 Q_x^2 Q_z^2 + K_3 Q_z^4}{\eta_a Q_x^2 + \eta_b Q_z^2}
\end{equation}
where $\eta_b = 2 \alpha_2 + \eta_{twist} + \eta_c$. In the same limit, Eq.~(7) yields
\begin{equation}
\Gamma_{n_1,n_2}^- = \frac{(K_1 Q_x^2 + K_2 Q_y^2)Q_\perp^2}{\eta_{splay} Q_x^2 + \eta_{twist} Q_y^2}
\end{equation} 
for the slower ($n_1$, $n_2$) mode in {\it Geometry 2}. The two Miesowicz viscosities $\eta_b$ and $\eta_a$ in Eq.~(8) correspond to sliding motions of smectic-$Z_A$ layers along the average director or normal to it, exactly as expected for the layer undulations coupled to director bend. Eq.~(9) reflects a coupling of splay and twist fluctuations particular to the smectic-$Z_A$ phase where the average $\hat{n}$ is parallel to the layers.

The solid blue curves in Fig.~7 are fits to Eq.~(8) for $\Gamma_{n_2,u}^-$ after substituting Eqs.~(1) for the dependence of $\vec{Q}$ on $\sin \theta_s$ and utilizing measured values of the refractive indices $n_\perp$, $n_\parallel$. They give a reasonably good description the data for the faster of the two modes detected experimentally in {\it Geometry 1}. Two parameters, $B \delta^2 / \eta_b$ and $\eta_a/\eta_b$, were varied at each temperature, while the parameter $K_3/\eta_b$ was constrained to the value for $K_3 / \eta_{bend}$ determined from Fig.~5, and $K_u$, $K_2$ were both set to zero (the latter contributing very weakly for small $Q_z$ and $Q_x$).

\begin{table}
	\caption{Parameter values obtained from the modeling of the relaxation rates of layer-director fluctuations in the smectic-$Z_A$ phase (Fig.~7).}
	\begin{ruledtabular}
		\begin{tabular}{cccc}
			$T-T_{NZ}$ ($^\circ$C) & $B \delta^2/K_3$ (m$^{-2}$) & $K_3$ (N) & $B \delta^2$ (N m$^{-2}$)\\
			\hline
			$-1.5$ & $6.1 \times 10^{13}$ & $7.3 \times 10^{-12}$ & $4.5 \times 10^2$ \\
			$-5.5$ & $14 \times 10^{13}$ & $7.5 \times 10^{-12}$ & $10 \times 10^2$ \\
			$-7.5$ & $16 \times 10^{13}$ & $7.7 \times 10^{-12}$ & $13\times 10^2$ \\
			$-11.5$ & $5.7 \times 10^{13}$ & $7.9 \times 10^{-12}$ & $4.5 \times 10^2$\\
		\end{tabular}
	\end{ruledtabular}
\end{table}

Table I summarizes results for the ratio $B \delta^2 / K_3$ at four temperatures spanning the smectic-$Z_A$ range. When combined with values for the bend elastic constant, $K_3$, in Fig.~2, these ratios yield $B \delta^2$ (last column of the table). Taking the chevron angle $\delta \approx 10^\circ$ in the middle of the smectic-$Z_A$ phase, as indicated by X-ray diffraction measurements \cite{Chen2023}, gives $B = 4.2 \times 10^4$~N\,m$^{-2}$, which is $\sim 100$ times smaller than in a typical calamitic smectic-$A$ liquid crystal \cite{Rouillon1990}. 
 
The variation of $B \delta^2$ with temperature in Table I could be due to an increase in the chevron angle $\delta$ below $T_{NZ}$, followed by a decrease approaching the transition to the ferroelectric nematic phase. This would also account for the dip in $I_s^{-1}$ observed for both the low and high values of $\theta_s$ studied in {\it Geometry 2} (Fig.~2), where the average director is normal to the scattering plane. The reason is that, when $\delta$ is nonzero, the scattering selection rules admit greater depolarized scattering from the fluctuating component of $\hat{n}$ parallel to the layers -- splay for small $\theta_s$ or twist for large $\theta_s$ -- which do not perturb the layer spacing or orientation. The enhancement of $I_s$ from these fluctuations in both cases is proportional to $\delta^2$. Thus, a dip and local minimum in $I_s^{-1}$ would arise from an increase in $\delta$, followed by a decrease, as the temperature decreases through the smectic-$Z_A$ phase.

Turning to the results in Fig.~8 for {\it Geometry 2}, the dashed blue curves are fits to Eq.~(9) for $\Gamma_{n_1,n_2}^-$ with three parameters, $\frac{K_1}{\eta_{splay}}$, $\frac{K_2}{ \eta_{splay}}$, and $\frac{\eta_{twist}}{\eta_{splay}}$ varied at each temperature. Although fairly accurate over the middle and upper range of $\sin^2 \theta_s$, the model systematically undershoots the data for $\sin^2\theta_s \lesssim 0.1$, especially at lower temperature. 

However, the description for small $\sin^2\theta_s$ can be improved if we consider the possibility of an anisotropic splay energy density. Up to now, we have not accounted for the electrostatic energy associated with splay of the polarization field and the corresponding accumulation of polarization charge, $\rho_P = -\vec{\nabla} \cdot \delta \vec{P} = -P_0 \vec{\nabla} \cdot \delta \vec{n}$, whose Fourier transform gives $-P_0 q_\perp n_1$. In the ferroelectric nematic phase, the effective splay energy density incorporating the electrostatic contribution is \cite{Eremin2022} $K_1^{eff} q_\perp^2 n_1^2 = \left(K_1 + \frac{P_0^2}{\epsilon_0 \epsilon_\perp (q_\perp^2 + \xi^{-2})} \right) q_\perp^2 n_1^2$, where $\epsilon_\perp$ is the dielectric constant for electric displacement normal to the director and $\xi$ is the characteristic (Debye) length for screening of $\rho_P$ by a finite concentration of mobile impurity ions. In the smectic-$Z_A$ phase, the average polarization is uniform in the directions ($\hat{x}$, $\hat{z}$) parallel to the layers but alternates along the layer normal ($\hat{y}$ direction). Therefore, for long wavelength splay ($q_\perp \ll \frac{2 \pi}{d_0}$), $\rho_P$ varies slowly in the $\hat{x}$ direction but reverses sign rapidly along $\hat{y}$. 

We therefore consider the possibility of an anisotropic screening length that varies in the $x$-$y$ plane as $\xi(\phi)^2 = \xi_x^2 \sin^2 \phi + \xi_y^2 \cos^2 \phi = \xi_y^2 +(\xi_x^2 - \xi_y^2) \sin^2 \phi$, where $\phi$ is the angle in the $x$-$y$ plane measured from the layer normal. Substituting this into the above expression for $K_1^{eff} q_\perp^2$, assuming $q_\perp \ll \xi (\phi)^{-2}$ (reasonable for optical $\vec{q}$), and noting $\sin \phi = \frac{q_x}{q_\perp}$, we obtain an anisotropic form for the effective splay elastic energy density, $\left[ \left( K_1 + \frac{P_0^2 \xi_y^2}{\epsilon_0 \epsilon_\perp} \right) q_\perp^2 + \frac{P_0^2 (\xi_x^2 - \xi_y^2)}{\epsilon_0 \epsilon_\perp} q_x^2 \right] n_1^2 \equiv ( K_1 q_\perp^2 + K_1^\prime q_x^2 ) n_1^2$. In the last expression on the right, a contribution due to polarization splay has been absorbed into a redefinition of $K_1$ and $K_1^\prime = \frac{P_0^2 (\xi_x^2 - \xi_y^2)}{\epsilon_0 \epsilon_\perp}$. Replacing $K_1 Q_\perp^2$ in Eq.~(9) with $K_1 Q_\perp^2 + K_1^\prime Q_x^2$ then yields
\begin{equation}
\Gamma_{n_1,n_2}^- = \frac{(K_1 Q_x^2 + K_2 Q_y^2)Q_\perp^2 + K_1^\prime Q_x^4}{\eta_{splay} Q_x^2 + \eta_{twist} Q_y^2}
\end{equation}

The solid blue lines in Fig.~8 are fits to Eq.~(10). Although these fits involve an additional variable parameter ($\frac{K_1^\prime}{\eta_{splay}}$), they better describe the data at low $\sin^2 \theta_s$. The ratio $\frac{K_1^\prime}{K_1}$ increases from $0.77$ to $4.6$ as $T-T_{NZ}$ decreases from $-2.5$ to $-11.5^\circ$C. This implies that, at the lowest temperature, the effective elastic constant for director splay fluctuations with $q = q_x$ (parallel to the smectic-$Z_A$ layers) is enhanced by a factor of $\sim 5-6$ relative to splay normal to the layers. The fits also give $\frac{K_1}{K_2} = 3-5$ in the middle of the $Z_A$ phase.

The modeling of $\Gamma_{n_1,n_2}^-$ in Fig.~8, with or without an anisotropic splay elasticity, also indicates a small ratio of twist to splay viscosity in the smectic-$Z_A$ phase. Specifically, we find $\frac{\eta_{twist}}{\eta_{splay}} \approx 0.1$, whereas the standard hydrodynamic theory for a uniaxial fluid predicts $\frac{\eta_{twist}}{\eta_{splay}} \gtrsim 1$ \cite{deGennes}. The inconsistency suggests, not surprisingly, that the nonpolar uniaxial form for the dissipative stress is incomplete. Perhaps inclusion of terms allowed by the polarity of the layers (wherein $\hat{n} \rightarrow -\hat{n}$ symmetry is broken), or by their biaxiality, could allow for an exceptional value of $\eta_{splay}$.

Finally, we fit the relaxation rate of the slow mode detected in {\it Geometry 1} (orange points in Fig.~7) to Eq.~(4) for $\Gamma_{n_1}$ with $K_1 Q_x^2$ replaced by $(K_1 + K_1^\prime) Q_x^2$ and using the expression $\eta_1 (\vec{Q}) = \eta_{bend} + \frac{(\eta{splay} - \eta_{bend}) Q_x^2}{Q_x^2 + a Q_z^2}$ to describe the crossover in the viscosity for $n_1$ fluctuations from $\eta_{splay}$ (when $Q_z \rightarrow 0$) to $\eta_{bend}$ (when $Q_x \rightarrow 0$). This expression is a simplification of the standard form for uniaxial splay-bend director fluctuations but has the virtue of adding only one extra parameter, $a$.

The solid orange lines in Fig.~7 are fits of the slower relaxation rate to the resulting form for $\Gamma_{n_1}$ with variable parameters $a$, $\frac{\eta_{bend}}{\eta_{splay}}$ and $\frac{K_1+K_1^\prime}{\eta_{splay}}$, $\frac{K_3}{\eta_{bend}}$ fixed to values from the analysis of data for $\Gamma_{n_2,u}^-$ and $\Gamma_{n_1,n_2}^-$ described above. For $\sin^2 \theta_s \gtrsim 0.1$ the models for $\Gamma_{n_2,u}^-$ and $\Gamma_{n_1}$ (blue and orange lines in Fig.~7) are essentially degenerate, which is consistent with the light scattering correlation function being described by a single exponential decay in this regime. Additionally, closer to the transitions to the para- and ferronematic phases, the relaxation rate for $\theta_s$ at the lower end of the single exponential regime better matches the model for $\Gamma_{n_2,u}^-$ than for $\Gamma_{n_1}$. This is exactly what one would expect if the chevron angle decreases near these transitions \cite{Chen2023}, thereby reducing the cross-section for scattering from $n_1$ fluctuations.

\section{Summary and Conclusion}

We reported on a dynamic light scattering study on the paraelectric nematic and anti-feroelectric smectic-$Z_A$ phases of the liquid crystal DIO. With decreasing temperature in the nematic phase, we observe a softening of the splay elastic constant followed by an increase just above the transition to the $Z_A$ phase. The twist and bend elasticities slowly increase with decreasing temperature, exhibiting no remarkable pretransitional behavior. On the other hand, both the splay and twist viscosities show very large pretransitional enhancements. 

The weak and continuous temperature dependence of the bend elasticity and viscosity at the transition to and within the smectic-$Z_A$ phase confirms that the layers form parallel to axis of molecular orientational order (director) -- a distinguishing feature of the $Z_A$ layer structure. 

We analyzed the dispersion of fluctuation relaxation rates in the $Z_A$ phase using a model based on an extension of the elastic free energy for a smectic-$C$ phase to the case of 90$^\circ$ director tilt, the approximation of the dissipative stress by the form derived for an incompressible, uniaxial fluid, and the incorporation of the effect of a chevron layer structure that develops in the $Z_A$ phase due to temperature-dependent layer shrinkage. The analysis yields an estimate of the elastic constant associated with layer compression that is two orders of magnitude lower than in an ordinary smectic-$A$ liquid crystal.

Although the model provides a reasonably accurate description of the fluctuation modes involving layer displacement $u$ and director $\hat{n}$, it does not account for an unusually high viscosity associated with splay fluctuations of $\hat{n}$ or for the possibility of an anisotropic splay constant, which seems to be indicated from the quantitative modeling of the relaxation rates. We suggested that the latter may arise from anisotropic screening of splay-induced polarization charge -- a point that needs further investigation and justification. In addition, it would be interesting to perform comparative light scattering studies of the director modes in other compounds and mixtures that possess an intermediate phase between the para- and ferroelectric nematic states, whether identified as the smectic-$Z_A$ phase or of an as yet undetermined nature.

\appendix

\section{Hydrodynamic equations}

We consider the smectic-$Z_A$ to be characterized by a one-dimensional density wave along $\hat{y}$, neglect permeation of molecules between layers, and assume the the system is incompressible. After spatial Fourier transformation, the hydrodynamic equations may be written as \cite{Orsay,deGennes}
\begin{equation}
\rho \frac{d\vec{v}}{dt} = i \vec{q} \cdot \overset{\text{\tiny$\leftrightarrow$}}{\sigma}^\prime - \frac{\delta f_{ZA}}{\delta u} - i\left( \vec{q} \cdot \overset{\text{\tiny$\leftrightarrow$}}{\sigma}^\prime \cdot \vec{q} \right) \frac{\vec{q}}{q^2} + q_y \frac{\delta f_{Z}}{\delta u} \frac{\vec{q}}{q^2}
\end{equation}
\begin{equation}
\frac{\delta f_{Z}}{\delta n_\alpha} = -h_\alpha~~(\alpha = x,y)
\end{equation}
with
\begin{equation}
\frac{\partial u}{\partial t} = v_y
\end{equation}
and
\begin{equation}
\vec{q} \cdot \vec{v} = 0
\end{equation}
Here $\rho$ is the mass density, $\vec{v}$, $u$, and $n_\alpha$ are the Fourier transforms of the velocity, layer displacement, and director fields, respectively; $\overset{\text{\tiny$\leftrightarrow$}}{\sigma}^\prime$ is the Fourier transform of the dissipative stress tensor; $f_{Z}$ is given by Eq.~(2) in the text; and $\vec{h}$ is the Fourier transform of the ``molecular field" or force conjugate to rate of change of the director relative to the background fluid. In the hydrodynamic theory, the components $\sigma_{\alpha \beta}^\prime$ and $h_\alpha$ are linear combinations of the velocity gradients $\partial_\gamma v_\delta$ and the time derivatives of the director components, $\frac{d n_\gamma}{dt}$. After Fourier transform, these components may be expressed in the $n_1$, $n_2$ basis defined in sec.~IVA. The last two terms on the right side of Eq.~(A1) enforce the incompressibility condition, Eq.~(A4).
  
To obtain Eqs.~(3), (4), and (7) of the main text, we drop the inertial term $\rho \frac{d \vec{v}}{dt}$ on the left side of Eq.~(A1), as is appropriate to describe the overdamped fluctuations measured experimentally, and we assume  the forms of $\overset{\text{\tiny$\leftrightarrow$}}{\sigma}^\prime$ and $\vec{h}$ from the standard hydrodynamic theory for a uniaxial fluid \cite{Leslie} are applicable (i.e, biaxiality of the smectic-$Z_A$ phase can be neglected).

\begin{acknowledgments}
The research reported here was supported by the National Science Foundation under grants no. DMR-2210083 (AG, PG, RT, JG, AJ, SS) and DMR-2341830 (BB, HW, OL).
\end{acknowledgments}
	
\bibliography{SmZA}

\onecolumngrid
\clearpage
\begin{center}
	\textbf{\large Supplementary Figures}
\end{center}
\setcounter{equation}{0}
\setcounter{figure}{0}
\setcounter{table}{0}
\setcounter{page}{1}
\makeatletter
\renewcommand{\theequation}{S\arabic{equation}}
\renewcommand{\thefigure}{S\arabic{figure}}
\renewcommand{\bibnumfmt}[1]{[S#1]}
\renewcommand{\citenumfont}[1]{S#1}

\vspace{-4cm}

\begin{figure}[H]
	%[h!]
	\centering
	%\captionsetup{font={small,stretch=0}}
	%\captionsetup{belowskip=-10pt}
	%\includegraphics[width=1\textwidth,trim={2.9cm 1.1cm 2.9cm 2.3cm},clip]{fig3.eps}
	\includegraphics[width=.8\textwidth]{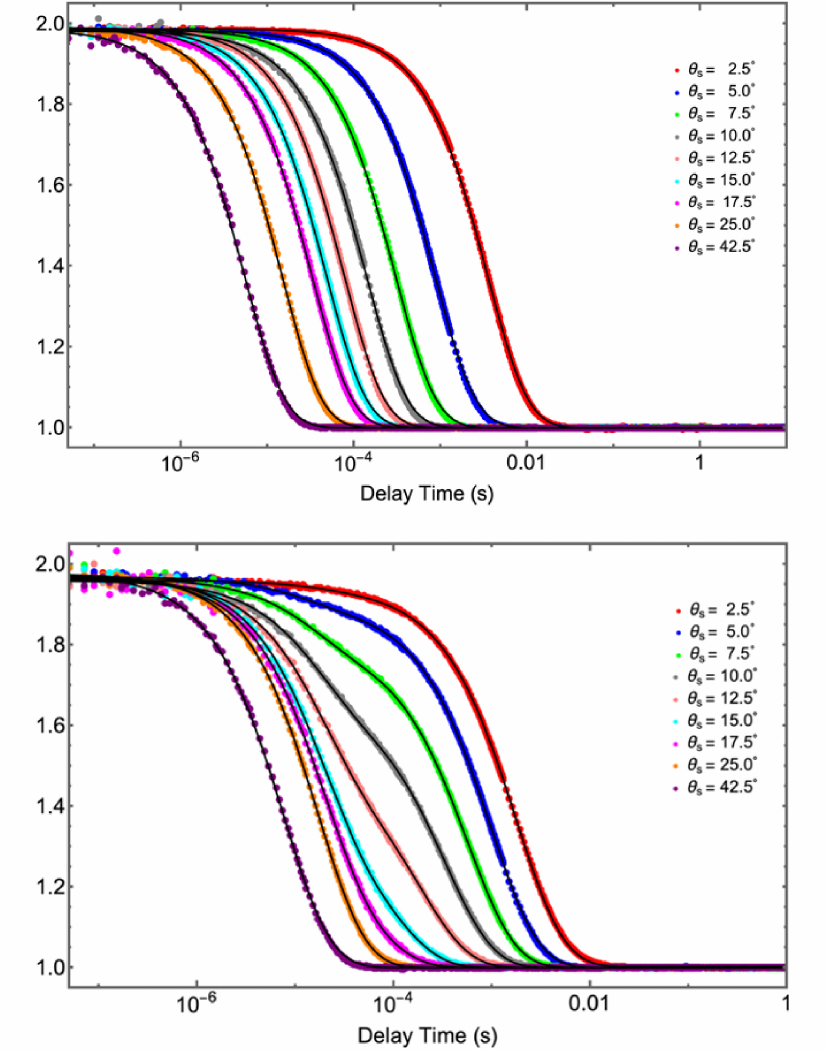}
	\caption{Normalized time correlation functions of the scattered light intensity acquired at various scattering angles in {\it Geometry 1} (see Fig.~1 of the main text). Top panel: Data in the nematic phase at $T-T_{NZ} = 6.5^\circ$C. The solid lines represent fits to a single exponential decay. Bottom panel: Data in the smectic-$Z_A$ phase at $T-T_{NZ} = -7.5^\circ$C. The fits for scattering angles in the range $2.5^\circ < \theta_s < 17.5^\circ$ correspond to a double exponential decay.}
\end{figure}

\pagebreak

\begin{figure}[H]
	%[h!]
	\centering
	%\captionsetup{font={small,stretch=0}}
	%\captionsetup{belowskip=-10pt}
	%\includegraphics[width=1\textwidth,trim={2.9cm 1.1cm 2.9cm 2.3cm},clip]{fig3.eps}
	\includegraphics[width=.8\textwidth]{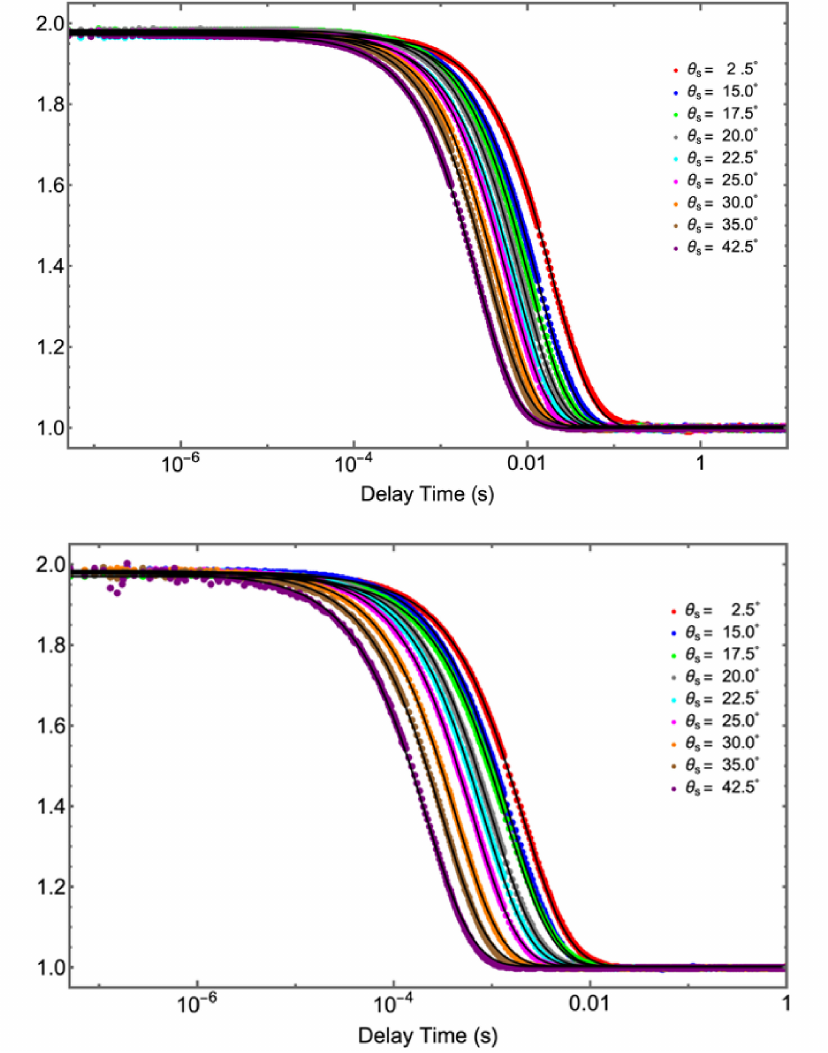}
	\caption{Normalized time correlation functions of the scattered light intensity acquired at various scattering angles in {\it Geometry 2} (see Fig.~1 of the main text). Top panel: Data in the nematic phase at $T-T_{NZ} = 6.5^\circ$C. Bottom panel: Data in the smectic-$Z_A$ phase at $T-T_{NZ} = -7.5^\circ$C. In both cases, the solid lines represent fits to a single exponential decay.}
\end{figure}

\pagebreak

\begin{figure}[H]
	%[h!]
	\centering
	%\captionsetup{font={small,stretch=0}}
	%\captionsetup{belowskip=-10pt}
	%\includegraphics[width=1\textwidth,trim={2.9cm 1.1cm 2.9cm 2.3cm},clip]{fig3.eps}
	\includegraphics[width=.6\textwidth]{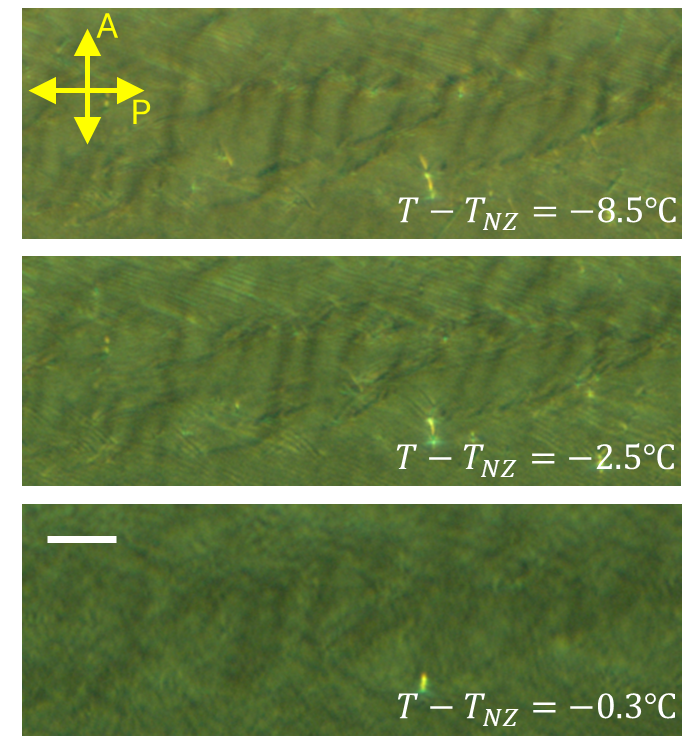}
	\caption{Optical texture of a band of zig-zag defects at various temperatures in the smectic-$Z_A$ phase of the $20~\mu$m thick DIO sample studied in this work. The texture is characteristic of chevron layer structure, as previously described in smectic-$Z_A$ samples (see refs.~19 and 20 of the main text). The planar-aligned director is rotated by $\sim 5^\circ$ with respect to the polarizer axis ($P$). Similar bands were observed in a parallel arrangement over the full ($2$~mm) field imaged by the microscope mounted on our light scattering set-up. Scale bar: $100~\mu$m. }
\end{figure}

\end{document}